\documentclass{article} 
\usepackage{iclr2023_conference,times}

\usepackage{amsmath,amsfonts,bm}

\def\eqref#1{equation~\ref{#1}}

\def\1{\bm{1}}

\DeclareMathAlphabet{\mathsfit}{\encodingdefault}{\sfdefault}{m}{sl}
\SetMathAlphabet{\mathsfit}{bold}{\encodingdefault}{\sfdefault}{bx}{n}

\newcommand{\E}{\mathbb{E}}

\newcommand{\Var}{\mathrm{Var}}

\usepackage{hyperref}
\usepackage{url}
 \usepackage[ruled]{algorithm2e}

\usepackage{graphicx}
\usepackage{capt-of}
\usepackage{caption}
\usepackage{float}
\usepackage{subfig}
\usepackage{amssymb}

\usepackage{threeparttable}
\usepackage{wrapfig}       
\usepackage{diagbox}
\usepackage{multirow}
\usepackage{makecell}
\usepackage{booktabs}       
\usepackage{pifont}

\usepackage{amsmath}
\usepackage{amsthm}
\newtheorem{theorem}{Theorem}

\newtheorem{proposition}{Proposition}

\def \P {\mathbb{P}}
\def \V {\Var}
\def \bfE {\mathbb{E}}

\def \cL{\mathcal{L}}

 \def \cD{\mathcal{D}}

\title{TDR-CL: Targeted Doubly Robust Collaborative Learning for Debiased Recommendations}

\author{Haoxuan Li$^1$\qquad Yan Lyu$^1$\qquad Chunyuan Zheng$^2$\qquad Peng Wu$^3$\thanks{Corresponding author.} \\
     $^1$Peking University\\
   $^2$University of California, San Diego \\
  $^3$Beijing Technology and Business University\\
     \texttt{\{hxli, lyuyan\}@stu.pku.edu.cn},  \texttt{czheng@ucsd.edu},  \texttt{pengwu@btbu.edu.cn}
}
\iclrfinalcopy 
\begin{document}

\maketitle


\begin{abstract}
Bias is a common problem inherent in recommender systems, which is entangled with users' preferences and poses a great challenge to unbiased learning. For debiasing tasks, the doubly robust (DR) method and its variants show superior performance due to the double robustness property, that is, DR is unbiased when either imputed errors or learned propensities are accurate.
However, our theoretical analysis reveals that DR usually has a large variance. Meanwhile, DR would suffer unexpectedly large bias and poor generalization caused by inaccurate imputed errors and learned propensities, which usually occur in practice.   
In this paper, we propose a principled approach that can effectively reduce the bias and variance \emph{simultaneously} for existing DR approaches when the error  imputation model is misspecified. 
In addition, we further propose a novel semi-parametric collaborative  learning approach that decomposes imputed errors into parametric and nonparametric parts and updates them collaboratively, resulting in more accurate predictions. Both theoretical analysis and experiments demonstrate the superiority of the proposed methods compared with existing debiasing methods.


\end{abstract}

\section{Introduction}

Addressing various tasks in recommender systems (RSs) with causality-based methods has become increasingly popular~\citep{Wu-etal2022}. 
 Causality-based recommendation 
 has shown its great potential in both numeric experiments and theoretical analyses across extensive literature~\citep{Chen-etal2020, Wang-Zhang-Sun-Qi2019}.  
Generally, the basic question faced in RS is that 
"what would the feedback be if recommending an item to a user", requiring to estimate the causal effect of a recommendation on user feedback. 
 To answer the question, many methods have been proposed,  
 such as  inverse propensity score (IPS)~\citep{Schnabel-Swaminathan2016}, self-normalized inverse propensity score (SNIPS)~\citep{ Swaminathan-Joachims2015}, error imputation based (EIB) methods~\citep{Steck2010}, and doubly robust (DR) methods~\citep{Chen-etal2021, Wang-Zhang-Sun-Qi2019, Wang-etal2021, Dai-etal2022, Ding-etal2022}. Among them, the DR method and its variants show  superior performance.  
 We compare and evaluate these methods in terms of three desired properties, including doubly robust~\citep{Hernan-Robins2020, Wu-Tan2022}, robust to small propensities~\citep{Rosenbaum2020},  
  and low variance~\citep{Tan-2007}. 
Failing to meet any of them may lead to sub-optimal performance~\citep{Molenberghs-etal2015, Laan-Rose2011}. 
Our theoretical analysis shows that DR has much greater variance and is less robust to small propensities compared to EIB~\citep{Kang-Schafer-2007}, even though the imputed errors and the learned propensities are accurate. Meanwhile, DR would suffer unexpectedly large bias and poor generalization caused by inaccurate imputed errors and learned propensities, which usually occur in practice.

In this paper, we first propose a novel targeted doubly robust (TDR) method, that can capture the merits of both DR and EIB effectively, by leveraging the targeted learning technique~\citep{Laan-Rose2011, Laan-Rose2018}. 
TDR can effectively reduce the bias and variance \emph{simultaneously} for existing DR approaches when the imputed errors are less accurate. Remarkably, TDR provides a model-agnostic framework and can be assembled into any DR method by updating its error  imputation model, resulting in more accurate predictions.  

To further reduce the bias and variance during the training process, we propose a novel uniform-data-free TDR-based collaborative learning (TDR-CL) approach that decomposes imputed errors into a parametric imputation model part and a nonparametric error part, where the latter adaptively rectifies the residual bias of the former. By updating the two parts collaboratively, TDR-CL achieves a more accurate and robust prediction.  Both theoretical analysis and experiments demonstrate the 
superiority of TDR and TDR-CL compared with existing methods.  


\section{Preliminaries} 
Many debiasing tasks in RS can be formulated using the widely adopted potential outcome framework~\citep{Neyman1990, Rubin1974}. 
Denote $\mathcal{U}=\{u\}$, $\mathcal{I}=\{i\}$ and $\mathcal{D} = \mathcal{U} \times \mathcal{I}$ as the sets of users, items and user-item pairs, respectively.  
  Let $x_{u,i}$, $r_{u,i}$, and $o_{u,i}$ be the feature, feedback, and  
  exposure status of user-item pair $(u,i)$, 
  where $o_{u,i} = 1$ or 0 represents whether the item $i$ is exposed to user $u$ or not. 
  Define $r_{u,i}(1)$  as the potential outcome if $o_{u,i}$ had been set to $1$, which is observed only when $o_{u,i} = 1$.     
 In RS, we are often interested in answering the causal question: "if we recommend products to users, what would be the feedback?". This question can be formulated as 
 to learn  
 the quantity $\E\left(r_{u,i}(1)|x_{u,i}\right)$, i.e., it requires to predict $r_{u,i}(1)$ using feature $x_{u,i}$, where $\bfE$ denotes the expectation with respect to the target distribution $\P$. Many classical tasks in RS  can be defined as estimating this quantity, such as rating prediction~\citep{Schnabel-Swaminathan2016} and   post-click conversion rate prediction~\citep{MRDR}. More examples can be found in \cite{Wu-etal2022}. 
  
Let $f_{\theta}(x_{u,i})$ be a model used to predict  $r_{u,i}(1)$  with parameter $\theta$. Ideally, 
if all $r_{u,i}(1)$ for $(u,i)\in \cD$ were observed,  $\theta$ can be trained directly by optimizing the following ideal loss 
    \[   \cL_{ideal} = |\cD|^{-1} \sum_{(u,i)\in \cD} e_{u,i},     \] 
where $e_{u,i}$ is the prediction error, e.g., the squared loss $e_{u,i} = (r_{u,i}(1) - f_{\theta}(x_{u,i})  )^2$. However, 
 since $r_{u,i}(1)$ is observed only when $o_{u,i} = 1$, the ideal loss is non-computable.  
   Restricting the analysis to non-missing data will  
  obtain biased conclusions, as the observed data may form an unrepresentative sample of the target population. 
 Different debiasing methods are designed to approximate and substitute  
  the ideal loss.  
For example, the IPS and EIB estimators  are given as 
         \[ \cL_{IPS} = |\cD|^{-1} \sum_{(u,i)\in \cD} o_{u,i} e_{u,i}/\hat p_{u,i}, \quad \cL_{EIB} = |\cD|^{-1} \sum_{(u,i)\in \cD} [ o_{u,i} e_{u,i} + (1-o_{u,i}) \hat e_{u,i}], \] 
where $\hat p_{u,i}$ is an estimate of propensity score $p_{u,i} := \P(o_{u,i}=1 | x_{u,i})$, $\hat e_{u,i}$ is  an estimate of prediction error $g_{u,i} := \bfE[ e_{u,i} | x_{u,i}]$, i.e., it fits $e_{u,i}$ using $x_{u,i}$.   
The DR estimator is formulated as 
    \[
        \cL_{DR} = |\cD|^{-1} \sum_{(u,i) \in \cD} \Big [ \hat e_{u,i}  +  \frac{ o_{u,i} (e_{u,i} -  \hat e_{u,i}) }{ \hat p_{u, i} } \Big ], \label{eq2} 
    \]
    which enjoys doubly robust property, i.e., it is an unbiased estimator of ideal loss when either imputed errors or learned propensities are accurate. 

\section{Motivation}


DR approaches have been extensively studied in RS for various debiasing tasks for its double robustness, e.g., rating prediction~\citep{Wang-Zhang-Sun-Qi2019, Wang-Liang-Charlin-Blei2020, Balance2023, SDR}, learning-to-rank (LTR)~\citep{saito2020doubly, oosterhuis2022doubly}, and post-click conversion rate prediction~\citep{MRDR, Dai-etal2022}, etc.
However, the DR still have several limitations that need to resolved. 
 We first show that DR has a large variance and is sensitive to small propensities as shown in Proposition \ref{th1} (see Appendix \ref{app-1} for  proofs).

\begin{proposition}
\label{th1} If $\hat p_{u,i}$ and $\hat e_{u,i}$ are accurate estimates of  $p_{u,i}$ and $g_{u,i}$, respectively, i.e., $\hat p_{u,i} = p_{u,i}$, $\hat e_{u,i} = g_{u,i}$,  then IPS, EIB and DR estimators are 
unbiased, and their variances satisfy 
			\begin{equation*}   \V( \cL_{EIB})  \leq  \V( \cL_{DR} ) \leq  \V( \cL_{IPS} ),     
			\end{equation*}
 where the equality holds if and only if $p_{u,i}= 1$ for all $(u,i) \in \cD$.  In addition, when $p_{u,i}$ tends to 0,  $\V( \cL_{IPS} )$ and  $\V( \cL_{DR} )$ tends to infinity, and $\V( \cL_{EIB})$ tends to its minimum.  
\end{proposition}

 Proposition \ref{th1} shows that the EIB estimator is low-variance and robust to small propensities~\citep{Tan-2007, Imbens-Rubin2015,  Wu-etal-2021, Wu-Han2022}. 
In RS, some small propensities will appear inevitably due to the sparsity of the exposed data, resulting in a significant difference between $\V( \cL_{EIB})$ and $\V( \cL_{DR})$.  
Nevertheless, EIB usually has a large bias  and is not preferred in practice. 
Proposition \ref{th1} provides a motivation to develop an estimator that combines the low-variance and robustness to small propensities of EIB with the double robustness of DR.    

 In summary, DR outperforms IPS in terms of both bias and variance. When compared with EIB, if $\hat e_{u,i}$ is inaccurate  but $\hat p_{u,i} $ is accurate, DR tends to have a smaller bias, but  if both $\hat e_{u,i}$ and $\hat p_{u,i}$ are accurate, then EIB has a smaller variance. 
 If $\hat e_{u,i}$ is accurate but $\hat p_{u,i} $ is inaccurate, then EIB  may be superior to DR  in terms of both bias and variance. 
In practice, both $\hat p_{u,i}$ and $\hat e_{u,i}$ are likely to be at least mildly inaccurate, so choosing from EIB and DR involves the bias-variance trade-off.  
 Ideally, it is desirable to develop a method that is robust to small propensities, with lower bias and variance compared to previous DR methods, 
 while maintaining the double robustness.
\section{Collaborative Learning Debiasing Framework} 
\subsection{Targeted Doubly Robust Estimator}    \label{sec-TMLE}

We first bridge the explicit form of the DR estimator and the EIB estimator by noting that
  \begin{equation}  \label{eq4}
        \cL_{DR}  
            ={} \underbrace{\frac{1}{|\cD|} \sum_{(u,i)\in \cD} [ o_{u,i} e_{u,i} + (1-o_{u,i}) \hat e_{u,i}]}_{\cL_{EIB}}  + \underbrace{\frac{1}{|\cD|} \sum_{(u,i)\in  \cD} o_{u,i}( e_{u,i} - \hat e_{u,i}) \frac{ 1 - \hat p_{u,i} }{\hat p_{u,i} }}_{\text{correction term}},   
    \end{equation}  
where $\cL_{DR}$ is formally equivalent to adding a correction term using learned propensities to $\cL_{EIB}$. The correction term has an important role in the bias-variance trade-off for the estimations of the ideal loss as shown in Proposition \ref{th1}. Specifically, compared with $\cL_{EIB}$, $\cL_{DR}$ can reduce bias by adding the correction term. 
As a compromise, the correction term will increase the variance of the DR estimator. 
Thus, if $\hat e_{u,i}$ is computed in a manner that ensures that 
\begin{equation}   \label{eq5}
		  \frac{1}{|\cD|} \sum_{(u,i)\in  \cD} o_{u,i}( e_{u,i} - \hat e_{u,i}) \frac{ 1 - \hat p_{u,i} }{\hat p_{u,i} }	  = 0.
		\end{equation}		
 then the EIB  estimator would have small bias and 
 the DR estimator would have small variance.   



For equation (\ref{eq5}) to hold, a naive method is taking it as a constraint condition when training the error  imputation model. However, the constraint (\ref{eq5}) may degrade the accuracy of the imputed errors because it will restrict the hypothesis space of the error  imputation model. 
  Instead of directly estimating $\hat e_{u,i}$ satisfying the constraint (\ref{eq5}), we propose to exploit the extra information on propensities when training the error imputation model.  
The basic idea of the proposed TDR estimator consists of the following two steps.

{\bf Step 1 (Initialization).} Let 
$\hat e_{u,i}$ 
be the imputed error obtained by using any of the existing DR methods.

{\bf Step 2 (Targeting).} 
Update $\hat e_{u,i}$ by fitting an extended one-parameter model as follows
	\begin{equation} \label{tmle}
		\tilde e_{u,i}(\eta) =   \hat e_{u,i} + \eta (1/ \hat p_{u,i} - 1 )  
	\end{equation} 
which includes a single variable $1/\hat p_{u,i}-1$ and the offset $\hat h(x_{u,i})$. The parameter $\eta$ is solved by minimizing the squared loss  between $\tilde e_{u,i}(\eta)$ and $e_{u,i}$ in the exposed events. 
Then the proposed TDR estimator is given as 
	\[    \cL_{TDR} = |\cD|^{-1} \sum_{(u,i) \in \cD} \Big [  \tilde e_{u,i}  +   o_{u,i} (e_{u,i} -  \tilde e_{u,i}) / \hat p_{u, i} \Big ].  \]
 The targeting step enlarges the hypothesis space of $\tilde  e_{u,i}$ compared to $\hat e_{u,i}$, and does not sacrifice the accuracy of the error  imputation model, due to the introduce of an error correction term $1/\hat p_{u,i}-1$ to estimate $e_{u,i}$. Theorem \ref{th2} shows the validity and preservation of TDR (see Appendix \ref{app-tmp}  for proofs). 


\begin{theorem} \label{th2}   
The imputed error $\tilde e_{u,i}$ obtained with TDR satisfies the following properties: \\
 (a) (validity) $\tilde e_{u,i}$ satisfies equation (\ref{eq5}), which implies TDR would have smaller bias than EIB and smaller variance than DR based on the initial imputed error $\hat e_{u,i}$. \\
(b) (preservation) $\hat \eta$ in the targeting step will converge to 0 and renders $\tilde e_{u,i} = \hat e_{u,i}$ when $\hat e_{u,i}$ already satisfies equation (\ref{eq5}).
\end{theorem}
{From Theorem \ref{th2}, TDR} guarantees that equation (\ref{eq5}) always holds, regardless of the choice of the initial imputed errors.
In addition, TDR inherits the desirable properties of EIB, such as low-variance and robust to small propensities, since equations (\ref{eq4}) and (\ref{eq5}) implies that the TDR estimator can be regarded as an EIB estimator. 

TDR would reduce the variance of DR as shown in Theorem \ref{th2}, a further question is whether the variance-reduction will come at the expense of an increase in bias? Remarkably, TDR has no sacrifice of bias. Specifically, 
it can be shown (see Appendix \ref{app-proof2}) that the bias of both $\cL_{DR}$ and $\cL_{TDR}$  are  
composed of the product of the errors of the propensity  model and imputation model weighted by $1/\hat p_{u,i}$. Therefore, given the same learned propensities, the more accurate the imputed errors are, the smaller the bias is.  
Since TDR updates $\hat e_{u,i}$  by adding an extra term $1/ \hat p_{u,i}-1$, so 
 $\tilde e_{u,i}$ is expected to be more accurate than $\hat e_{u,i}$, resulting in a smaller bias for $\cL_{TDR}$ than $\cL_{DR}$. 

 Importantly, the TDR provides a model-agnostic framework due to the free choice of the initial imputed errors in Step 1, which has great potentially strengths for recommendation. TDR can be assembled into any competing DR approach~\citep{Wang-Zhang-Sun-Qi2019, MRDR, Dai-etal2022}, by updating its error imputation model with the targeting step. This extra targeting step tends to reduce both the bias and variance of the competing DR approach, resulting in more accurate predictions. 
Next, Theorem \ref{th3} indicates the double robustness of the TDR estimator (see Appendix \ref{app-proof2} for proofs). 



\begin{theorem} \label{th3} 
The proposed TDR estimator have the following properties:   \\
(a) (unbiasedness under accurate imputed errors) $\cL_{TDR}$ is unbiased if $\tilde e_{u,i}$ accurately estimates $g_{u,i}$.\\ 
(b) (unbiasedness under accurate learned propensities) Suppose that $\hat p_{u,i}$ accurate estimates $p_{u,i}$, and the validity of $\hat e_{u,i}$ doesn't hold, then  $\cL_{EIB}$ is biased, while $\cL_{TDR}$ is unbiased.
\end{theorem}

Besides, Theorem \ref{th3}(b) reveals that TDR can remove the bias of $\cL_{EIB}$  even though the initial imputed errors are inaccurate, provided the learned propensities are accurate.   

\subsection{Semi-parametric Collaborative Learning} 

In this subsection, we propose a novel {TDR}-based collaborative  learning ({TDR-CL}) approach, in which the imputed errors $\tilde e_{u,i}$ are decomposed into a parametric error imputation model part $\hat e_{u,i}$ and a nonparametric targeting part $\omega_{u, i}\triangleq \eta (1/ \hat p_{u,i} - 1 )$ as in Section \ref{sec-TMLE}, 
i.e., $\tilde e_{u,i}=\hat e_{u,i}+\omega_{u,i}$. The latter corrects the residuals of the error  imputation model. By updating both the parametric and nonparametric parts collaboratively, the bias and variance of the TDR estimator can be further reduced, resulting in more accurate predictions.

First, the embedding of each user $u$ and item $i$ is obtained by matrix factorization, and the stack layer gets the embedding $x_{u,i}$ by concatenation. TDR-based learning methods require estimated propensities for all user-item pairs, thus the Naive Bayes approach is no longer applicable. To handle this problem, the pre-trained propensities are obtained by conducting logistic regression of $o_{u, i}$ on $x_{u,i}$, and the model parameters are used as the initialization of $p_{\xi}(x_{u,i})$ in the iterative learning process.  
Given both the parametric error imputation  part $\hat e_{u, i}=g_\phi(x_{u,i})$ and nonparametric targeting part $\omega_{u, i}$, the propensity model $p_\xi(x_{u,i})$ and the prediction model $f_\theta(x_{u,i})$ are updated simultaneously using the training loss
\begin{equation*}
\cL_{{TDR-CL}}\left(\theta, \xi, \phi \right)
=  \cL_{TDR} + |\cD|^{-1} \sum_{ (u, i) \in \mathcal{D}} \Big[ -
o_{u, i} \cdot \log \hat p_{u,i}-(1-o_{u, i}) \cdot \log (1-\hat p_{u,i})\Big], 
\end{equation*}  
where $\hat p_{u,i}=p_\xi(x_{u,i})$, $e_{u, i}=(f_{\theta}(x_{u, i})-r_{u, i}(1))^{2}$,  
$\tilde e_{u, i}=(f_{\theta}\left(x_{u, i}\right)-g_{\phi}\left(x_{u, i}\right)-{\omega_{u, i}} - \perp (f_{\theta}(x_{u, i})))^{2}$ with $\perp$ the operator that sets the gradient of the operand to zero thus $\nabla_{\theta} \perp\left(f_{\theta}\left(x_{u, i}\right)\right)=0$ and $\perp\left(f_{\theta}\left(x_{u, i}\right)\right)=f_{\theta}\left(x_{u, i}\right)$.

Then, unlike traditional alternative learning algorithms that directly use the parametric part $g_{\phi}\left({x}_{u, i}\right)$ as  {$\tilde {e}_{u, i}$}, the proposed collaborative  learning additionally uses $\omega_{u,i}$ as a non-parametric correction term 
summed with $g_{ \phi}\left({x}_{u, i}\right)$ to correct the  estimation of ${e}_{u, i}$. Specifically, given the prediction model and the propensity model, {$\tilde e_{u,i}$} first updates its parametric part $g_{ \phi}\left({x}_{u, i}\right)$ by minimizing \begin{equation*}
\mathcal{L}_{\mathrm{e}}(\theta, \xi, \phi)=|\cD|^{-1} \sum_{ (u, i) \in \cD} o_{u,i} \left( \tilde{e}_{u, i}-e_{u, i}\right)^{2} / \hat{p}_{u, i},     
\end{equation*}
where $e_{u, i}=r_{u, i}(1)-f_{\theta}\left(x_{u, i}\right), {\tilde e_{u, i}=g_{\phi}\left(x_{u, i}\right)+\omega_{u,i}}$. Next, the targeting step described in Section \ref{sec-TMLE} is applied 
to further update the imputed errors {$\tilde e_{u,i}$}.  Through calculating the optimal step size for line search $\eta^{\ast} = \arg\min\limits_{\eta}\sum o_{u,i}(e_{u,i}(\theta)-\hat{e}_{u,i}(\phi)- \eta(1/\hat{p}_{u,i}-1))^2$, the non parametric {targeted} error term $\omega_{u,i}$ is updated by adding $\eta^{\ast}(1/\hat{p}_{u,i}-1)$. 

In summary, the proposed learning approach
collaboratively update the parametric term $\hat e_{u, i}=g_{\phi}\left(x_{u, i}\right)$ and the nonparametric term $\omega_{u,i}$ to achieve a better trade-off to estimate $e_{u,i}$, which can reduce the bias of the existing DR methods such as DR-JL~\citep{Wang-Zhang-Sun-Qi2019} and MRDR-DL~\citep{MRDR}, by further modeling for the fitted residuals of the parametric parts $\hat e_{u, i}$. On the other hand, as shown in Theorems \ref{th2} and \ref{th3}, when $\hat e_{u, i}$ is already an accurate estimate of $e_{u,i}$, the introduction of the {targeted error term $\tilde{e}_{u, i}$} satisfies no-harm property and the unbiasedness is maintained. We summarized the proposed {TDR-CL}  approach in Alg. \ref{alg1}.

\begin{algorithm}[t]
\caption{The Proposed {Targeted Doubly Robust Collaborative Learning, TDR-CL}}
\label{alg1}
\LinesNumbered 
\KwIn{{observed ratings $\mathbf{R}^{o}$, pre-trained learned propensities $ \mathbf{\hat P}$, and {$\omega_{u,i}= 0$.}}}
\While{stopping criteria is not satisfied}{
    \For{number of steps for training the prediction and propensity model}{Sample a batch of user-item pairs $\left\{\left(u_{j}, i_{j}\right)\right\}_{j=1}^{J}$ from $\mathcal{D}$\;
    Update $\theta$ and $\xi$ by descending along the gradient $\nabla_{\theta, \xi} \cL_{{TDR-CL}}\left(\theta, \xi, \phi \right)$\;
    }
    \For{number of steps for training the imputation model with targeting step}{Sample a batch of user-item pairs $\left\{\left(u_{k}, i_{k}\right)\right\}_{k=1}^{K}$ from $\mathcal{O}$\;
    Update $\phi$ by descending along the gradient $\nabla_{\phi} \mathcal{L}_{\mathrm{e}}(\theta, \xi, \phi)$\;
    Sample a batch of user-item pairs $\left\{\left(u_{l}, i_{l}\right)\right\}_{l=1}^{L}$ from $\mathcal{D}$\;
    $\eta^{\ast} \gets \arg\min\limits_{\eta}\sum o_{u,i}(e_{u,i}(\theta)-\hat{e}_{u,i}(\phi)-\eta({1}/{\hat{p}_{u,i}}-1))^2$\;
    Update $\omega_{u,i} \gets \omega_{u,i}+\eta^{\ast}({1}/{\hat{p}_{u,i}}-1)$ for all user-item pairs.} 
}
\end{algorithm}

\section{Semi-synthetic Experiments}
 In this section, following the previous studies~\citep{Schnabel-Swaminathan2016, Wang-Zhang-Sun-Qi2019, saito2020doubly, MRDR}, we aim to answer the following research question (RQ) on the semi-synthetic datasets:
	\begin{enumerate}  
	\item[\bf RQ1.]  Does the proposed {TDR} estimator in estimating the ideal loss have both the statistical properties of lower bias and variance in the presence of  selection bias?
	\end{enumerate}

\subsection{Experimental Setup}
{\bf Dataset and Preprocessing.} \textbf{MovieLens 100K\footnote{https://grouplens.org/datasets/movielens/100k/} (ML-100K)} is a dataset of 100,000 missing-not-at-random (MNAR) ratings from 943 users and 1,682 movies collected from movie recommendation ratings. \textbf{MovieLens 1M\footnote{https://grouplens.org/datasets/movielens/1m/} (ML-1M)} is a larger dataset of 1,000,209 MNAR ratings from 6,040 users and 3,952 movies. Following the data preprocessing procedure of previous studies~\citep{Schnabel-Swaminathan2016, Wang-Zhang-Sun-Qi2019, saito2020doubly, MRDR},  we first use matrix factorization \citep{koren2009matrix} to complete the rating matrix in the five-scale. Then for each predicted ratings $R_{u, i} \in\{1,2,3,4,5\}$, we assign the $p_{u, i} \in(0,1)$ with $p_{u, i}=p \alpha^{\max \left(1,5-R_{u, i}\right)}$. Finally, we replace the predicted ratings $R_{u, i}$ with $r_{u, i}^{\text {true}}$ in $\{0.1,0.3,0.5,0.7,0.9\}$ and sample the binary click indicator and conversion label with the Bernoulli sampling
$
o_{u, i} \sim \operatorname{Bern}(p_{u, i}), r_{u, i} \sim \operatorname{Bern}(r_{u, i}^{true}), \forall(u, i) \in \mathcal{D}, 
$
where $\operatorname{Bern}(\cdot)$ denotes the Bernoulli distribution. 

{\bf Predicted Metrics.} The following prediction metrics are used to evaluate the debiasing performance under different scenarios.

\vspace{-4pt}
$\bullet$ \textbf{ONE:} $\hat{r}_{u, i}$ is identical to the $r_{u, i}^{\text {true}}$, except that $|\{(u, i) \mid r_{u, i}^{\text {true}}=0.9\}|$ randomly selected $r_{u, i}^{\text {true}}$ of $0.1$ are flipped to 0.9.

\vspace{-4pt}
$\bullet$ \textbf{THREE:} Same as ONE, but flipping $r_{u, i}^{\text {true}}$ of $0.3$ instead.

\vspace{-4pt}    
$\bullet$ \textbf{FIVE:} Same as ONE, but flipping $r_{u, i}^{\text {true}}$ of $0.5$ instead.

\vspace{-4pt}    
$\bullet$ \textbf{ROTATE:} $\hat{r}_{u, i}=r_{u, i}-0.2$ when $r_{u, i} \geq 0.3$, and $\hat{r}_{u, i}=0.9$ when $r_{u, i}=0.1$.

\vspace{-4pt}    
$\bullet$ \textbf{SKEW:} $\hat{r}_{u, i}$ follows the truncated Gaussian distribution $\mathcal{N}_{[0.1,0.9]}(\mu=r_{u, i}^{\text{true}}, \sigma= (1-r_{u, i}^{\text {true }})/{2})$.

\vspace{-4pt}    
$\bullet$     \textbf{CRS:} $\hat{r}_{u, i}=0.2$ if the $r_{u, i}^{\text {true}} \leq 0.6$. Otherwise, $\hat{r}_{u, i}=0.6$.

{\bf Experimental Details.} For each prediction matrix $\mathbf{\hat R} = \{\hat r_{u,i}(1): (u,i)\in\cD\}$, the proposed TDR is compared with  Naive~\citep{koren2009matrix}, EIB~\citep{Lobato-etal2014, Steck2010}, IPS~\citep{saito2020ips, Schnabel-Swaminathan2016}, and DR~\citep{Wang-Zhang-Sun-Qi2019, saito2020doubly} methods. 
We obtain the propensities by ${1}/{\hat{p}_{u, i}}=(1-\beta)/{p_{u, i}}+\beta/{p_{e}}$, where $p_{e}= |\mathcal{D}|^{-1} \sum_{(u, i) \in \mathcal{D}} o_{u, i}$, and $\beta$ is randomly sampled from $[0, 1]$ to introduce noises. Define $\hat e_{u,i} = \mathrm{CE}( \sum_{(u, i) \in \mathcal{O}} r_{u, i} w_{u,i} , \hat{r}_{u, i}),$ where
$w_{u,i} = \left( 1/ \hat{p}_{u, i} \right) \big / ( \sum_{(u, i) \in \mathcal{O}} {1 / \hat{p}_{u, i}})$, $\mathrm{CE}$ denotes the cross entropy loss. 
 For EIB and DR, the imputed error is computed as $\tilde{e}_{u, i}= \hat e_{u,i}$, 
 For TDR, 
 $\tilde{e}_{u,  i}= \hat e_{u,i} + \eta^{\ast}({1}/{\hat{p}_{u,i}}-1)$, 
 where $\eta^{\ast}=\arg\min\limits_{\eta}\sum_{(u, i) \in \mathcal{O}}(e_{u,i}- \hat e_{u,i}-\eta({1}/{\hat{p}_{u,i}}-1))^2$.  
 The performance of the estimators is based on the absolute relative error ($\mathrm{RE}$) of the estimated and ideal loss
 $\mathrm{RE} (\mathcal{L}_{{est}})= |\mathcal{L}_{{ideal}}(\hat{\mathbf{R}})-\mathcal{L}_{ {est}}(\hat{\mathbf{R}})| / \mathcal{L}_{{ideal}}(\hat{\mathbf{R}})$,  
where $\cL_{{est}}$ denotes the estimator to be compared. $\mathrm{RE}$ evaluates the accuracy of the estimated loss, and a smaller $\mathrm{RE}$ value indicates a higher estimation accuracy.

\subsection{Experiment Results (RQ1)}
In Table \ref{Semi}, we report the means and standard deviations of the RE of the five estimators for each predicted matrix over 20 times of sampling. On the one hand, the average RE of the IPS, DR and TDR methods is significantly lower than that of the Naive method, verifying the validity of causal-based debiasing methods.  The proposed TDR achieves the lowest RE in all settings, attributed to the introduced correction term $\omega_{u, i}$ for estimating $ e_{u,i}$, that further reduces the bias of DR. The direct application of the EIB method is even worse than the Naive method, attributed to the challenge to make an accurate estimate of $ e_{u,i}$. On the other hand, same as the conclusion of Theorem \ref{th1}, the standard deviation of the EIB method is significantly lower compared to the IPS and DR methods. The proposed TDR method combines the advantages of the EIB in terms of lower standard deviations than IPS and DR in all settings, reflecting stronger robustness. It can be concluded that the estimation accuracy and robustness of the proposed method are significantly improved compared to the previous methods.

\begin{table}[t]
 \centering
  \scriptsize
 \setlength{\tabcolsep}{2.5pt}
\captionof{table}{Mean and standard deviation of the relative error on the Naive, EIB, IPS, DR and TDR.} 
\vspace{-0.3cm}
\begin{tabular}{c|c|cccccc}
\toprule
Dataset                    & {Method} & ONE                  & THREE                  & FIVE       & ROTATE           & SKEW               & CRS             
\\ \midrule
\multicolumn{1}{c|}{}      & Naive                          & 0.0688 $\pm$ 0.0025 & 0.0790 $\pm$ 0.0028 & 0.1027 $\pm$ 0.0028  & 0.1378 $\pm$ 0.0011 & 0.0265 $\pm$ 0.0021 & 0.1062 $\pm$ 0.0022\\
                           & EIB                            & 0.5442 $\pm$ 0.0016 & 0.5878 $\pm$ 0.0017 & 0.6167 $\pm$ 0.0018 & 0.2533 $\pm$ 0.0004 & 0.3584 $\pm$ 0.0007 & 0.1443 $\pm$ 0.0007        \\
\multicolumn{1}{c|}{ML-100K}  & IPS                     & 0.0338 $\pm$ 0.0033 & 0.0390 $\pm$ 0.0037 & 0.0511 $\pm$ 0.0033 & 0.0696 $\pm$ 0.0026 & 0.0129 $\pm$ 0.0027 & 0.0526 $\pm$ 0.0026 \\ 
                           & DR                        & 0.0140 $\pm$ 0.0034 & 0.0180 $\pm$ 0.0037 & 0.0150 $\pm$ 0.0034 & 0.0401 $\pm$ 0.0016 & 0.0101 $\pm$ 0.0027 & 0.0237 $\pm$ 0.0025 \\ 
                           & TDR                           & \textbf{0.0053 $\pm$ 0.0026*} & \textbf{0.0035 $\pm$ 0.0025*} & \textbf{0.0066 $\pm$ 0.0032*} & \textbf{0.0325 $\pm$ 0.0015*} & \textbf{0.0029 $\pm$ 0.0020*} &
                           \textbf{0.0193 $\pm$ 0.0025*}\\ \midrule
                           
                           & Naive                          & 0.0682 $\pm$ 0.0007          & 0.0783 $\pm$ 0.0007          & 0.1014 $\pm$ 0.0008                    & 0.1377 $\pm$ 0.0005           & 0.0256 $\pm$ 0.0007 & 0.1054 $\pm$ 0.0006           \\
                           & EIB                            & 0.5437 $\pm$ 0.0005          & 0.5872 $\pm$ 0.0005          & 0.6157 $\pm$ 0.0005          & 0.2531 $\pm$ 0.0001          & 0.3575 $\pm$ 0.0002 & 0.1442 $\pm$ 0.0001          \\
\multicolumn{1}{c|}{ML-1M} & IPS                     & 0.0343 $\pm$ 0.0009          & 0.0394 $\pm$ 0.0009          & 0.0508 $\pm$ 0.0009          & 0.0687 $\pm$ 0.0006          & 0.0130 $\pm$ 0.0008          & 0.0528 $\pm$ 0.0007          \\
                           & DR                        & 0.0130 $\pm$ 0.0009          & 0.0168 $\pm$ 0.0009          & 0.0133 $\pm$ 0.0009          & 0.0399 $\pm$ 0.0005          & 0.0090 $\pm$ 0.0008 & 0.0229 $\pm$ 0.0007  \\
                           & TDR                           & \textbf{0.0054 $\pm$ 0.0009*} & \textbf{0.0031 $\pm$ 0.0009*} & \textbf{0.0076 $\pm$ 0.0009*} & \textbf{0.0324 $\pm$ 0.0005*} & \textbf{0.0031 $\pm$ 0.0008*} & 
                           \textbf{0.0187 $\pm$ 0.0007*}\\ \bottomrule
\end{tabular}\label{Semi}
\begin{tablenotes}
\scriptsize
\item Note: * means statistically significant results ($\text{p-value} \leq 0.001$) using the paired-t-test compared with the best baseline.
\end{tablenotes}
\end{table}

\section{Real-world Experiments}
 In this section, we conduct experiments to evaluate the proposed methods on two real-world benchmark datasets containing missing-at-random (MAR) ratings. Throughout, our methods are implemented without uniform data to estimate the propensities, which differs from the existing Naive Bayes approach. We aim to answer the following RQs:
\begin{enumerate}
	\item[\bf RQ2.]  How do the proposed methods compare with the existing methods in terms of debiasing performance in practice?
	\item[\bf RQ3.]  How does the collaborative learning phase design affect the performance of our methods?  
	\item[\bf RQ4.]  Do our methods stably perform well under different learned propensities?
	\end{enumerate}
\subsection{Experimental Setup}
{\bf Dataset and Preprocessing.} MAR ratings are necessary to evaluate the performance of debiasing methods on real-world datasets. Following previous studies, we take the following two benchmark datasets: {\bf Coat Shopping\footnote{https://www.cs.cornell.edu/\textasciitilde schnabts/mnar/}} has 4,640 MAR and 6,960 MNAR ratings of 290 users to 300 Coats. {\bf Music! R3\footnote{http://webscope.sandbox.Music.com/}} has 54,000 MAR and 311,704 MNAR ratings of 15,400 users to 1,000 songs.

{\bf Baselines.} We take the widely used Matrix Factorization
(MF) as the base model~\citep{koren2009matrix}, and compare the proposed methods with the following baselines: Base Model~\citep{koren2009matrix}, IPS~\citep{Schnabel-Swaminathan2016}, SNIPS~\citep{Swaminathan-Joachims2015}, IPS with asymmetric training (IPS-AT)~\citep{saito2020asymmetric},  CVIB~\citep{wang2020information},
DIB~\citep{liu2021mitigating}, DR~\citep{saito2020doubly}, DR-JL~\citep{Wang-Zhang-Sun-Qi2019}, {DR-CL}, MRDR-JL~\citep{MRDR}, {MRDR-CL}, where {DR-CL and MRDR-CL} are performed using the proposed Alg. \ref{alg1}, but without the {targeting} step update (lines 9-11), also for comparison purpose. In addition, the proposed TDR-based methods include TDR, TDR-JL, and TMRDR-JL implemented by {a single targeting step}, 
 and {TDR-CL and TMRDR-CL} implemented by collaborative learning approach as shown in Alg. \ref{alg1}. The real-world experimental protocols and details are provided in Appendix \ref{app-real}.



\subsection{Performance Comparison (RQ2)}
In Table \ref{tab2}, we report the performance of various debiasing methods using MSE, AUC, NDCG@5, and NDCG@10 as evaluation metrics. {For previous de-biasing methods, propensity-based IPS, SNIPS, IPS-AT, and information bottleneck-based CVIB and DIB all outperform the base model. The doubly robust methods, such as DR-JL, DR-CL, MRDR-JL, and MRDR-CL, using alternating learning and outperforming DR, which are considered as the most competitive baselines.} The proposed {TDR} estimators are implemented by both {single-step} and collaborative learning, respectively, based on DR and MRDR as initialized error  imputation models, outperforming the baseline methods significantly on all AUC, NDCG@5, and NDCG@10 metrics, attributed to the effectiveness of the introduced nonparametric correction term. It is noted that the collaborative version of {TDR} achieves the optimal performance both within DR and MRDR, which implements the proposed targeting step repeatedly. The fact that TDR-JL and TMRDR-JL implemented the targeting step only at the final training of the prediction models outperformed DR-JL and MRDR-JL, respectively, further 
demonstrates the effectiveness of the proposed  targeting step to correct imputed errors. 

\begin{table}[t]
\centering
 \small
\setlength{\tabcolsep}{7.5pt}
\captionof{table}{MSE, AUC, NDCG@5, and NDCG@10 on the MAR test set of Coat and Music. We bold the outperforming DR-based and MRDR-based models. The proposed {TDR} methods implemented by {a  single targeting step} are marked with $\ast$ and collaborative learning are marked with $\dagger$.}
\begin{tabular}{lcccccccc}
\toprule
                      & \multicolumn{4}{c}{Coat}                                             & \multicolumn{4}{c}{Music}                                             \\ \cmidrule{2-9}
                      & MSE             & AUC             & N@5          & N@10         & MSE             & AUC             & N@5          & N@10         \\ \midrule \midrule
Base Model            & 0.2448          & 0.7047          & 0.5912          & 0.6667          & 0.2494          & 0.6795          & 0.6353          & 0.7644          \\ 
+ IPS                 & 0.2389          & 0.7041          & 0.6170          & 0.6852          & 0.2496          & 0.6824          & 0.6409          & 0.7674          \\
+ SNIPS               & 0.2388          & 0.7061          & 0.6145          & 0.6945          & 0.2493          & 0.6815          & 0.6454          & 0.7701          \\
+ IPS-AT     & 0.2344            & \textbf{0.7320} &  0.6102 & 0.6784 & 0.2480 & 0.6816 & 0.6409 & 0.7667 \\
+ CVIB  & \textbf{0.2201}          & 0.7234          & 0.6221          & 0.6991          & 0.2638          & 0.6823          & 0.6483          & 0.7719          \\
+ DIB   & 0.2334 & 0.7104 & \textbf{0.6303} & 0.6986  & 0.2494 & 0.6832 & 0.6348 & 0.7633  \\
\midrule
+ DR                  & 0.2359          & 0.7031          & 0.6213          & 0.6967          & \textbf{0.2420} & 0.6867          & 0.6613          & 0.7791          \\
+ DR-JL               & 0.2352          & 0.7155          & 0.6183          & 0.6925          & 0.2496          & 0.6853          & 0.6536          & 0.7738          \\
+ DR-CL               & 0.2358          & 0.7183          & 0.6261          & 0.6927          & 0.2494          & 0.6808          & 0.6334          & 0.7622          \\
\textbf{+ TDR$^\ast$}    & 0.2268 & 0.7109 & 0.6300 & \textbf{0.7006} & \textbf{0.2115} & \textbf{0.7044} & \textbf{0.7008} & \textbf{0.8016} \\
\textbf{+ TDR-JL$^\ast$}  & \textbf{0.2151} & \textbf{0.7236} & \textbf{0.6388} & \textbf{0.7047} & 0.2577          & \textbf{0.7036} & \textbf{0.6786} & \textbf{0.7884} \\
\textbf{+ TDR-CL$^\dagger$} & \textbf{0.2119} & \textbf{0.7339} & \textbf{0.6526} & \textbf{0.7112} & \textbf{0.2472} & \textbf{0.7057} & \textbf{0.6758} & \textbf{0.7871} \\ \midrule
+ MRDR-JL             & 0.2162          & 0.7192          & 0.6360          & 0.7016          & 0.2496 & 0.6842          & 0.6487          & 0.7717          \\
+ MRDR-CL             & 0.2155          & 0.7200          & 0.6427          & 0.7047          & \textbf{0.2494} & 0.6805          & 0.6345          & 0.7623          \\
\textbf{+ TMRDR-JL$^\ast$} & \textbf{0.2114} & \textbf{0.7278} & \textbf{0.6498} & \textbf{0.7101} & 0.2557          & \textbf{0.7036} & \textbf{0.6785} & \textbf{0.7884} \\
\textbf{+ TMRDR-CL$^\dagger$} & \textbf{0.2114} & \textbf{0.7316} & \textbf{0.6428} & \textbf{0.7088} & \textbf{0.2473}          & \textbf{0.7060} & \textbf{0.6803} & \textbf{0.7902} \\ \bottomrule
\end{tabular} \label{tab2}
\end{table}

\subsection{In-depth Analysis (RQ3, RQ4)}
\begin{figure}[t]
    \centering
\subfloat[Comparison of DR-JL, DR-CL and TDR-CL in terms of MSE, AUC and NDCG@5.]{
\begin{minipage}[t]{1\linewidth}
\centering
\includegraphics[width=0.98 \textwidth]{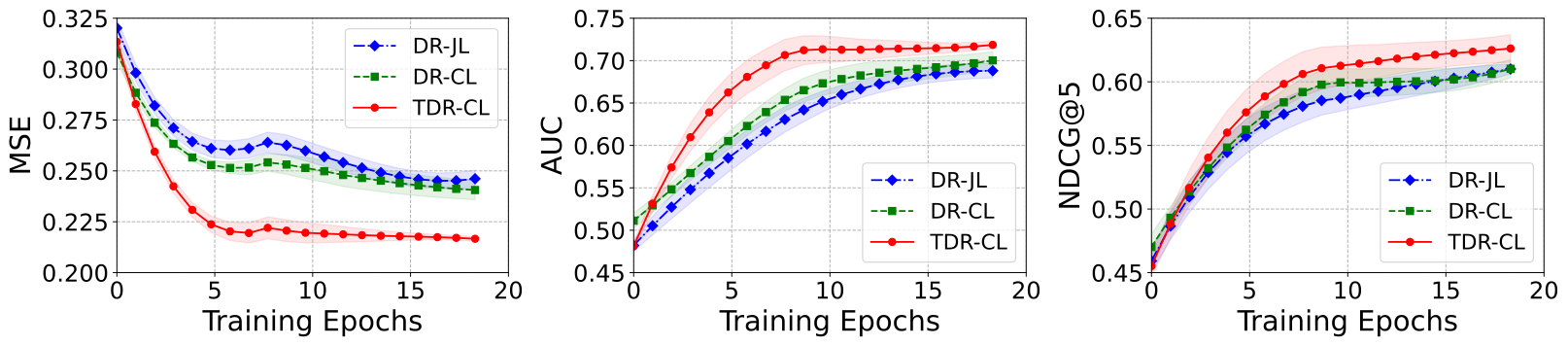}
\end{minipage}%
}%

\subfloat[Comparison of MRDR-JL, MRDR-CL and TMRDR-CL in terms of MSE, AUC and NDCG@5.]{
\begin{minipage}[t]{1\linewidth}
\centering
\includegraphics[width=0.98\textwidth]{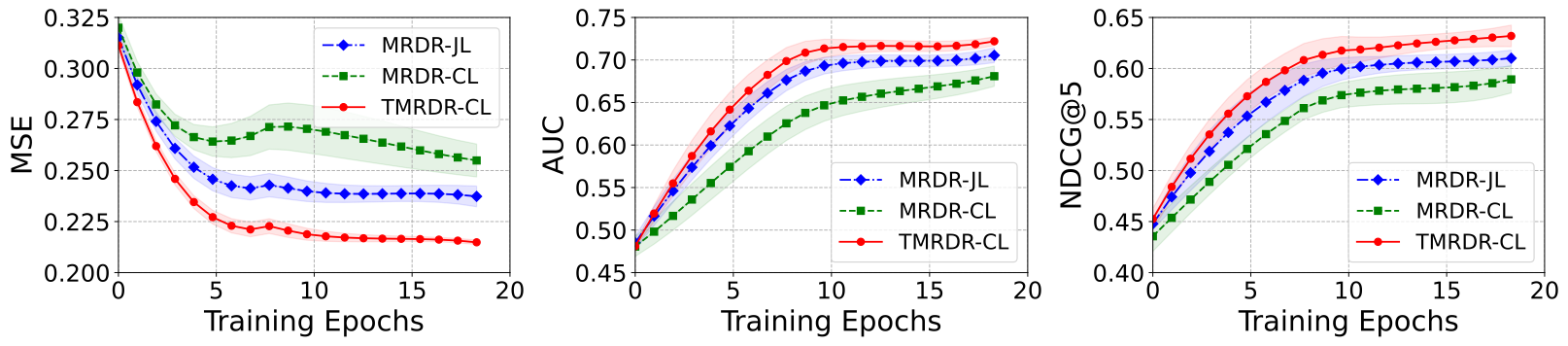}
\end{minipage}%
}%
\caption{Ablation studies on DR methods (top) and MRDR methods (bottom), where DR-CL and MRDR-CL skips the targeting steps in TDR-CL and TMRDR-CL.}
    \label{fig:opt} 
\end{figure}

{ \bf Ablation Study (RQ3).} 
To illustrate the specific reasons for the effectiveness of the TDR-CL algorithm, we conduct ablation studies on DR-based and MRDR-based methods, respectively. From Figure \ref{fig:opt}, DR-CL and DR-JL perform similarly on MSE, AUC and NDCG@5 metrics, and the MRDR approach has similar findings, which indicates that the directly use of collaborative learning approach without targeting steps cannot improve prediction performance. However, for the proposed TDR-CL and TMRDR-CL methods, there is a significant performance improvement compared to the DR-CL and MRDR-CL methods without targeting steps. This ablation study reveals that the improvement in the proposed TDR-CL and TMRDR-CL  originates from the nonparametric correction term of the imputed errors, not from introducing additional model parameters for updating.

\begin{figure}[t]
    \centering
    \includegraphics[scale=0.50]{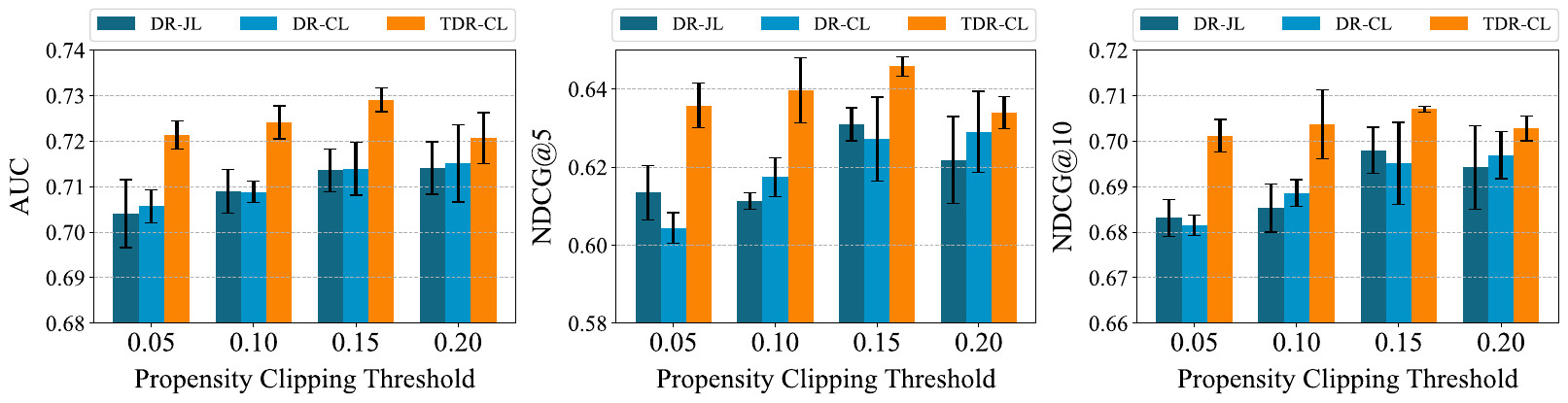}
    \caption{Learning performance on MAR test set of AUC (left), NDCG@5 (middle), and NDCG@10 (right) with varying levels of propensity clipping threshold.}
    \label{fig:sensitive} 
    \vskip -0.2in
\end{figure}

{\bf Effect on Learned Propensities (RQ4).} An important fact is that the nonparametric correction term in the {TDR} estimator is based on given learned propensities. In order to examine whether the proposed targeting steps stably help to improve the prediction accuracy under different learned propensities obtained by setting different clipping threshold, we conducted repeated experiments to quantify the sensitivity of the TDR-CL method to the propensity clipping threshold. From Figure \ref{fig:sensitive}, 
the proposed method outperforms the {DR-JL and DR-CL} methods in terms of AUC, NDCG@5, and NDCG@10 on all clipping thresholds. The optimal performance is reached when the clipping threshold is equal to 0.15, which is interpreted as achieving the best trade-off between information utilization and robustness.



\section{Related Work}

{\bf Debiasing in Recommendation.}  Bias is a common problem inherent in RS~\citep{Chen-etal2020, Wu-etal2022}, such as   
 popularity bias~\citep{zhang2021causal}, model selection bias~\citep{CIKM-YuanHYZCDL19}, user self-selection bias~\citep{saito2020asymmetric}, position bias~\citep{ai2018unbiased}, and conformity bias~\citep{DBLP:conf/recsys/LiuCY16}. Various methods were proposed for unbiased learning. 
 For example, \citet{Schnabel-Swaminathan2016} considered the recommendation as treatment and introduced the IPS and self-normalized IPS (SNIPS) methods to debiasing in explicit feedback data. \citet{saito2020ips}  
 extended it to the implicit recommendation. \citet{Wang-Zhang-Sun-Qi2019} proposed a doubly robust joint learning approach that improved the IPS method. Subsequently, 
 a series of enhanced DR methods were developed, such as MRDR~\citep{MRDR}, Multi-task DR~\citep{Zhang-etal2020}, DR-MSE~\citep{Dai-etal2022},  BRD-DR~\citep{Ding-etal2022}, and SDR~\citep{SDR}. \cite{MR2023} proposed a multiple robust method that takes the advantages of multiple propensity and error imputation models.      
 In addition, several new debiasing algorithm are designed via using an extra small uniform dataset~\citep{bonner2018causal, Chen-etal2021,liu2020general,Wang-etal2021, Balance2023}. \cite{Chen-etal2020}  provided a thorough discussion the recent progress on debiasing tasks in RS.  
 \cite{Wu-etal2022} established a unified causal analysis framework and gave formal causal definitions of various 
biases in RS from the perspective of causal inference. 
 Unlike the existing enhanced DR approaches that purse a better bias-variance trade-off, the proposed TDR reduces both the bias and variance and is theoretically guaranteed.

{\bf Targeted Learning.} Targeted learning is a general framework in causal inference~\citep{Laan-Rose2011} that includes many field-specific approaches to accommodate various scientific problems in different fields, 
such as survival analysis~\citep{Stitelman-etal2012}, genomics~\citep{Gruber-Laan2010}, epidemiology~\citep{Rose-Laan2014} and etc. More application scenarios about targeted learning can refer to the two excellent monographs~\citep{Laan-Rose2011, Laan-Rose2018}.  \cite{Dragon} proposed adapting neural networks for estimating the average treatment effects based on targeted learning.  
 Different from the existing literature of targeted learning, this paper deals with the estimator and learning problem simultaneously. 
 To the best of our knowledge, this is the first paper that extends targeted learning to the field of debiased recommendation. 
   \vspace{-8pt}
\section{Conclusion} 
   \vspace{-8pt}
In this paper, we propose a TDR estimator for debiased recommendation that enjoys the properties of double robustness, boundedness, low variance, and robustness to small propensities simultaneously. 
 Theoretical analysis shows that TDR can effectively reduce the bias and variance simultaneously for any DR estimator when the error  imputation model is less accurate. 
In addition, we further propose a novel uniform-data-free TDR-based  collaborative learning approach that adaptively implements the targeting step,  
thus making the prediction model more robust. We conducted experiments on both semi-synthetic and real-world data. The superiority of the proposed method is demonstrated when compared with the existing debiasing methods.  
Throughout, we adopt $1/\hat p_{u,i}-1$ as a key choice of targeting step to satisfy equation (\ref{eq5}), which can be regraded as a first-order targeted learning~\cite{carone2014higher}. In future work, we will explore higher-order targeted learning and more effective feature selection in the proposed targeting step. 


\section*{Ethics Statement} 
This work is mostly theoretical and experiments are based on synthetic and public datasets. We claim that this work does not present any foreseeable negative social impact.

\section*{Reproducibility Statement} 
Code is provided in Supplementary Materials to reproduce the experimental results.




\subsubsection*{Acknowledgments}
The work was supported by the National Key R\&D Program of China under Grant No. 2019YFB1705601. 

\bibliography{iclr2023_conference}

\begin{thebibliography}{48}
\providecommand{\natexlab}[1]{#1}
\providecommand{\url}[1]{\texttt{#1}}
\expandafter\ifx\csname urlstyle\endcsname\relax
  \providecommand{\doi}[1]{doi: #1}\else
  \providecommand{\doi}{doi: \begingroup \urlstyle{rm}\Url}\fi

\bibitem[Ai et~al.(2018)Ai, Bi, Luo, Guo, and Croft]{ai2018unbiased}
Qingyao Ai, Keping Bi, Cheng Luo, Jiafeng Guo, and W~Bruce Croft.
\newblock Unbiased learning to rank with unbiased propensity estimation.
\newblock In \emph{SIGIR}, 2018.

\bibitem[Bonner \& Vasile(2018)Bonner and Vasile]{bonner2018causal}
Stephen Bonner and Flavian Vasile.
\newblock Causal embeddings for recommendation.
\newblock In \emph{RecSys}, 2018.

\bibitem[Carone et~al.(2014)Carone, D{\'\i}az, and van~der
  Laan]{carone2014higher}
Marco Carone, Iv{\'a}n D{\'\i}az, and Mark~J van~der Laan.
\newblock Higher-order targeted minimum loss-based estimation.
\newblock \emph{U.C. Berkeley Division of Biostatistics Working Paper Series},
  Paper 331, 2014.

\bibitem[Chen et~al.(2020)Chen, Dong, Wang, Feng, Wang, and He]{Chen-etal2020}
Jiawei Chen, Hande Dong, Xiang Wang, Fuli Feng, Meng Wang, and Xiangnan He.
\newblock Bias and debias in recommender system: A survey and future
  directions.
\newblock \emph{arXiv:2010.03240}, 2020.

\bibitem[Chen et~al.(2021)Chen, Dong, Qiu, He, Xin, Chen, Lin, and
  Yang]{Chen-etal2021}
Jiawei Chen, Hande Dong, Yang Qiu, Xiangnan He, Xin Xin, Liang Chen, Guli Lin,
  and Keping Yang.
\newblock Autodebias: Learning to debias for recommendation.
\newblock In \emph{SIGIR}, 2021.

\bibitem[Dai et~al.(2022)Dai, Li, Wu, Dong, Zhou, Zhang, He, Zhang, and
  Sun]{Dai-etal2022}
Quanyu Dai, Haoxuan Li, Peng Wu, Zhenhua Dong, Xiao-Hua Zhou, Rui Zhang,
  Xiuqiang He, Rui Zhang, and Jie Sun.
\newblock A generalized doubly robust learning framework for debiasing
  post-click conversion rate prediction.
\newblock In \emph{KDD}, 2022.

\bibitem[Ding et~al.(2022)Ding, Wu, Feng, He, Wang, Liao, and
  Zhang]{Ding-etal2022}
Sihao Ding, Peng Wu, Fuli Feng, Xiangnan He, Yitong Wang, Yong Liao, and
  Yongdong Zhang.
\newblock Addressing unmeasured confounder for recommendation with sensitivity
  analysis.
\newblock In \emph{KDD}, 2022.

\bibitem[Gruber \& van~der Laan(2010)Gruber and van~der Laan]{Gruber-Laan2010}
Susan Gruber and Mark~J. van~der Laan.
\newblock An application of collaborative targeted maximum likelihood
  estimation in causal inference and genomics.
\newblock \emph{The International Journal of Biostatistics}, 6:\penalty0
  articel 18, 2010.

\bibitem[Guo et~al.(2021)Guo, Zou, Liu, Ye, Cheng, Wang, Chen, Yin, and
  Chang]{MRDR}
Siyuan Guo, Lixin Zou, Yiding Liu, Wenwen Ye, Suqi Cheng, Shuaiqiang Wang,
  Hechang Chen, Dawei Yin, and Yi~Chang.
\newblock Enhanced doubly robust learning for debiasing post-click conversion
  rate estimation.
\newblock In \emph{SIGIR}, 2021.

\bibitem[Hern{\'a}n \& Robins(2020)Hern{\'a}n and Robins]{Hernan-Robins2020}
Miguel~A. Hern{\'a}n and James~M. Robins.
\newblock \emph{Causal Inference: What If}.
\newblock Boca Raton: Chapman and Hall/CRC, 2020.

\bibitem[Hernández-Lobato et~al.(2014)Hernández-Lobato, Houlsby, and
  Ghahramani]{Lobato-etal2014}
José~Miguel Hernández-Lobato, Neil Houlsby, and Zoubin Ghahramani.
\newblock Probabilistic matrix factorization with non-random missing data.
\newblock In \emph{ICML}, 2014.

\bibitem[Imbens \& Rubin(2015)Imbens and Rubin]{Imbens-Rubin2015}
Guido~W. Imbens and Donald~B. Rubin.
\newblock \emph{Causal Inference For Statistics Social and Biomedical Science}.
\newblock Cambridge University Press, 2015.

\bibitem[Kang \& Schafer(2007)Kang and Schafer]{Kang-Schafer-2007}
Joseph~D.Y. Kang and Joseph~L. Schafer.
\newblock Demystifying double robustness: a comparison of alternative
  strategies for estimating a population mean from incomplete data.
\newblock \emph{Statistical Science}, 22:\penalty0 523--539, 2007.

\bibitem[Koren et~al.(2009)Koren, Bell, and Volinsky]{koren2009matrix}
Yehuda Koren, Robert Bell, and Chris Volinsky.
\newblock Matrix factorization techniques for recommender systems.
\newblock \emph{Computer}, 42\penalty0 (8):\penalty0 30--37, 2009.

\bibitem[Li et~al.(2023{\natexlab{a}})Li, Dai, Li, Lyu, Dong, Zhou, and
  Wu]{MR2023}
Haoxuan Li, Quanyu Dai, Yuru Li, Yan Lyu, Zhenhua Dong, Xiao-Hua Zhou, and Peng
  Wu.
\newblock Multiple robust learning for recommendation.
\newblock In \emph{AAAI}, 2023{\natexlab{a}}.

\bibitem[Li et~al.(2023{\natexlab{b}})Li, Xiao, Zheng, and Wu]{Balance2023}
Haoxuan Li, Yanghao Xiao, Chunyuan Zheng, and Peng Wu.
\newblock Balancing unobserved confounding with a few unbiased ratings in
  debiased recommendations.
\newblock In \emph{WWW}, 2023{\natexlab{b}}.

\bibitem[Li et~al.(2023{\natexlab{c}})Li, Zheng, and Wu]{SDR}
Haoxuan Li, Chunyuan Zheng, and Peng Wu.
\newblock Stable{D}{R}: Stabilized doubly robust learning for recommendation on
  data missing not at random.
\newblock In \emph{ICLR}, 2023{\natexlab{c}}.

\bibitem[Liu et~al.(2020)Liu, Cheng, Dong, He, Pan, and Ming]{liu2020general}
Dugang Liu, Pengxiang Cheng, Zhenhua Dong, Xiuqiang He, Weike Pan, and Zhong
  Ming.
\newblock A general knowledge distillation framework for counterfactual
  recommendation via uniform data.
\newblock In \emph{SIGIR}, 2020.

\bibitem[{Liu} et~al.(2021){Liu}, {Cheng}, {Zhu}, {Dong}, {He}, {Pan}, and
  {Ming}]{liu2021mitigating}
Dugang {Liu}, Pengxiang {Cheng}, Hong {Zhu}, Zhenhua {Dong}, Xiuqiang {He},
  Weike {Pan}, and Zhong {Ming}.
\newblock Mitigating confounding bias in recommendation via information
  bottleneck.
\newblock In \emph{RecSys}, 2021.

\bibitem[Liu et~al.(2016)Liu, Cao, and Yu]{DBLP:conf/recsys/LiuCY16}
Yiming Liu, Xuezhi Cao, and Yong Yu.
\newblock Are you influenced by others when rating?: Improve rating prediction
  by conformity modeling.
\newblock In \emph{RecSys}, 2016.

\bibitem[Molenberghs et~al.(2015)Molenberghs, Fitzmaurice, Kenward, Tsiatis,
  and Verbeke]{Molenberghs-etal2015}
Geert Molenberghs, Garrett Fitzmaurice, Michael~G. Kenward, Anastasios Tsiatis,
  and Geert Verbeke.
\newblock \emph{Handbook of Missing Data Methodology}.
\newblock Chapman \& Hall/CRC, 2015.

\bibitem[Neyman(1990)]{Neyman1990}
Jerzy~Splawa Neyman.
\newblock On the application of probability theory to agricultural experiments.
  essay on principles. section 9.
\newblock \emph{Statistical Science}, 5:\penalty0 465--472, 1990.

\bibitem[Oosterhuis(2022)]{oosterhuis2022doubly}
Harrie Oosterhuis.
\newblock Doubly-robust estimation for unbiased learning-to-rank from
  position-biased click feedback.
\newblock \emph{arXiv preprint arXiv:2203.17118}, 2022.

\bibitem[Rose \& van~der Laan(2014)Rose and van~der Laan]{Rose-Laan2014}
Sherri Rose and Mark~J. van~der Laan.
\newblock A double robust approach to causal effects in case-control studies.
\newblock \emph{American journal of epidemiology}, 179:\penalty0 663--669,
  2014.

\bibitem[Rosenbaum(2020)]{Rosenbaum2020}
Paul~R. Rosenbaum.
\newblock \emph{Design of Observational Studies}.
\newblock Springer Nature Switzerland AG, second edition, 2020.

\bibitem[Rubin(1974)]{Rubin1974}
Donald~B. Rubin.
\newblock Estimating causal effects of treatments in randomized and
  nonrandomized studies.
\newblock \emph{Journal of educational psychology}, 66:\penalty0 688--701,
  1974.

\bibitem[{Saito}(2020)]{saito2020asymmetric}
Yuta {Saito}.
\newblock Asymmetric tri-training for debiasing missing-not-at-random explicit
  feedback.
\newblock In \emph{SIGIR}, 2020.

\bibitem[Saito(2020)]{saito2020doubly}
Yuta Saito.
\newblock Doubly robust estimator for ranking metrics with post-click
  conversions.
\newblock In \emph{SIGIR}, 2020.

\bibitem[Saito et~al.(2020)Saito, Yaginuma, Nishino, Sakata, and
  Nakata]{saito2020ips}
Yuta Saito, Suguru Yaginuma, Yuta Nishino, Hayato Sakata, and Kazuhide Nakata.
\newblock Unbiased recommender learning from missing-not-at-random implicit
  feedback.
\newblock In \emph{WSDM}, 2020.

\bibitem[Schnabel et~al.(2016)Schnabel, Swaminathan, Singh, Chandak, and
  Joachims]{Schnabel-Swaminathan2016}
Tobias Schnabel, Adith Swaminathan, Ashudeep Singh, Navin Chandak, and Thorsten
  Joachims.
\newblock Recommendations as treatments: Debiasing learning and evaluation.
\newblock In \emph{ICML}, 2016.

\bibitem[Shi et~al.(2019)Shi, Blei, and Veitch]{Dragon}
Claudia Shi, David~M. Blei, and Victor Veitch.
\newblock Adapting neural networks for the estimation of treatment effects.
\newblock In \emph{NeurIPS}, 2019.

\bibitem[Steck(2010)]{Steck2010}
Harald Steck.
\newblock Training and testing of recommender systems on data missing not at
  random.
\newblock In \emph{KDD}, 2010.

\bibitem[Stitelman et~al.(2012)Stitelman, Gruttola, and van~der
  Laan]{Stitelman-etal2012}
Ori~M. Stitelman, Victor~De Gruttola, and Mark~J. van~der Laan.
\newblock A general implementation of tmle for longitudinal data applied to
  causal inference in survival analysis.
\newblock \emph{The International Journal of Biostatistics}, 8:\penalty0
  articel 26, 2012.

\bibitem[Swaminathan \& Joachims(2015)Swaminathan and
  Joachims]{Swaminathan-Joachims2015}
Adith Swaminathan and Thorsten Joachims.
\newblock The self-normalized estimator for counterfactual learning.
\newblock In \emph{NeurIPS}, 2015.

\bibitem[Tan(2007)]{Tan-2007}
Zhiqiang Tan.
\newblock Comment: understanding {O}{R}, {P}{S} and {D}{R}.
\newblock \emph{Statistical Science}, 22:\penalty0 560--568, 2007.

\bibitem[van~der Laan \& Rose(2011)van~der Laan and Rose]{Laan-Rose2011}
Mark~J. van~der Laan and Sherri Rose.
\newblock \emph{Targeted Learning: Causal Inference for Observational and
  Experimental Data}.
\newblock Springer, 2011.

\bibitem[van~der Laan \& Rose(2018)van~der Laan and Rose]{Laan-Rose2018}
Mark~J. van~der Laan and Sherri Rose.
\newblock \emph{Targeted Learning in Data Science: Causal Inference for Complex
  Longitudinal Studies}.
\newblock Springer, 2018.

\bibitem[Wang et~al.(2019)Wang, Zhang, Sun, and Qi]{Wang-Zhang-Sun-Qi2019}
Xiaojie Wang, Rui Zhang, Yu~Sun, and Jianzhong Qi.
\newblock Doubly robust joint learning for recommendation on data missing not
  at random.
\newblock In \emph{ICML}, 2019.

\bibitem[Wang et~al.(2021)Wang, Zhang, Sun, and Qi]{Wang-etal2021}
Xiaojie Wang, Rui Zhang, Yu~Sun, and Jianzhong Qi.
\newblock Combating selection biases in recommender systems with a few unbiased
  ratings.
\newblock In \emph{WSDM}, 2021.

\bibitem[Wang et~al.(2020{\natexlab{a}})Wang, Liang, Charlin, and
  Blei]{Wang-Liang-Charlin-Blei2020}
Yixin Wang, Dawen Liang, Laurent Charlin, and David~M. Blei.
\newblock Causal inference for recommender systems.
\newblock In \emph{RecSys}, 2020{\natexlab{a}}.

\bibitem[Wang et~al.(2020{\natexlab{b}})Wang, Chen, Wen, Huang, Kuruoglu, and
  Zheng]{wang2020information}
Zifeng Wang, Xi~Chen, Rui Wen, Shao-Lun Huang, Ercan~E. Kuruoglu, and Yefeng
  Zheng.
\newblock Information theoretic counterfactual learning from
  missing-not-at-random feedback.
\newblock In \emph{NeurIPS}, 2020{\natexlab{b}}.

\bibitem[Wu et~al.(2021)Wu, Xu, Tong, Jiang, and Lu]{Wu-etal-2021}
Peng Wu, Xinyi Xu, Xingwei Tong, Qing Jiang, and Bo~Lu.
\newblock Semiparametric estimation for average causal effects using propensity
  score-based spline.
\newblock \emph{Journal of Statistical Planning and Inference}, 212:\penalty0
  153--168, 2021.

\bibitem[Wu et~al.(2022{\natexlab{a}})Wu, Han, Tong, and Li]{Wu-Han2022}
Peng Wu, Shasha Han, Xingwei Tong, and Runze Li.
\newblock Propensity score regression for causal inference with treatment
  heterogeneity.
\newblock \emph{Statistica Sinica}, 2022{\natexlab{a}}.

\bibitem[Wu et~al.(2022{\natexlab{b}})Wu, Li, Deng, Hu, Dai, Dong, Sun, Zhang,
  and Zhou]{Wu-etal2022}
Peng Wu, Haoxuan Li, Yuhao Deng, Wenjie Hu, Quanyu Dai, Zhenhua Dong, Jie Sun,
  Rui Zhang, and Xiao-Hua Zhou.
\newblock On the opportunity of causal learning in recommendation systems:
  Foundation, estimation, prediction and challenges.
\newblock In \emph{IJCAI}, 2022{\natexlab{b}}.

\bibitem[Wu et~al.(2022{\natexlab{c}})Wu, Tan, Hu, and Zhou]{Wu-Tan2022}
Peng Wu, Zhiqiang Tan, Wenjie Hu, and Xiao-Hua Zhou.
\newblock Model-assisted inference for covariate-specific treatment effects
  with high-dimensional data.
\newblock \emph{Statistica Sinica}, 2022{\natexlab{c}}.

\bibitem[Yuan et~al.(2019)Yuan, Hsia, Yang, Zhu, Chang, Dong, and
  Lin]{CIKM-YuanHYZCDL19}
Bo{-}Wen Yuan, Jui{-}Yang Hsia, Meng{-}Yuan Yang, Hong Zhu, Chih{-}Yao Chang,
  Zhenhua Dong, and Chih{-}Jen Lin.
\newblock Improving ad click prediction by considering non-displayed events.
\newblock In \emph{CIKM}, 2019.

\bibitem[Zhang et~al.(2020)Zhang, Bao, Liu, Yang, Lin, Wen, and
  Ramezani]{Zhang-etal2020}
Wenhao Zhang, Wentian Bao, Xiao-Yang Liu, Keping Yang, Quan Lin, Hong Wen, and
  Ramin Ramezani.
\newblock Large-scale causal approaches to debiasing post-click conversion rate
  estimation with multi-task learning.
\newblock In \emph{WWW}, 2020.

\bibitem[Zhang et~al.(2021)Zhang, Feng, He, Wei, Song, Ling, and
  Zhang]{zhang2021causal}
Yang Zhang, Fuli Feng, Xiangnan He, Tianxin Wei, Chonggang Song, Guohui Ling,
  and Yongdong Zhang.
\newblock Causal intervention for leveraging popularity bias in recommendation.
\newblock In \emph{SIGIR}, 2021.

\end{thebibliography}
\bibliographystyle{iclr2023_conference}

\newpage 
\appendix

\section{Proof of Proposition 1}  \label{app-1} 
Recall that  $p_{u,i} =\P(o_{u,i}=1| x_{u,i}) = \bfE[o_{u,i}|x_{u,i}]$ and $g_{u,i}=\bfE[e_{u,i}| x_{u,i}]$, both of them are functions of $x_{u,i}$.  Throughout, we maintain the common unconfoundedness assumption (i.e., $r_{u,i}(1)\perp \!\!\! \perp o_{u,i}\mid x_{u,i}$) and the consistency assumption, (i.e., $r_{u,i}(1) = r_{u,i}$ if $o_{u,i} = 1$). All the lower-case letters denote random variables for simplification. 

\begin{proof}[Proof of Proposition 1] {The property of unbiasedness is obvious}.  Next, we focus on analysing the variance. Define 
    \begin{align*}
        \sigma^2(x_{u,i}) ={}& \V(e_{u,i} | x_{u,i}) =   \bfE[ (e_{u,i} - g_{u,i})^2 \mid x_{u,i} ], 
    \end{align*}
then $\bfE[ e_{u,i}^2 | x_{u,i}] = \sigma^2(x_{u,i}) +  g_{u,i}^2$. 
The variance of IPS estimator is given by 
    \begin{align*}
       \V( \cL_{IPS} ) 
        ={}&   |\cD|^{-1} \cdot   \V(  \frac{ o_{u,i} e_{u,i} }{ p_{u,i} }    ) \\
        ={}&  |\cD|^{-1} \cdot  \Biggl [  \bfE[\frac{ o_{u,i}^2 e_{u,i}^2 }{ p_{u,i}^2 }  ]  -  \Big \{ \bfE( \frac{ o_{u,i} e_{u,i} }{ p_{u,i} }  ) \Big \}^2 \Biggr ] \\
        ={}&  |\cD|^{-1} \cdot  \Biggl [  \bfE \left \{\frac{ \bfE[o_{u,i} | x_{u,i}] \cdot \bfE[e_{u,i}^2| x_{u,i}] }{ p_{u,i}^2 }  \right\}  -  \Big \{ \bfE\left ( \frac{ \bfE[o_{u,i} | x_{u,i}] \cdot \bfE[e_{u,i}|  x_{u,i}] }{ p_{u,i} } \right ) \Big \}^2 \Biggr ] \\  
        ={}&  |\cD|^{-1} \cdot  \Biggl [  \bfE[\frac{ e_{u,i}^2 }{ p_{u,i} }  ]  -  \Big \{ \bfE( e_{u,i}) \Big \}^2 \Biggr ] \\
            ={}&  |\cD|^{-1} \cdot  \Biggl [  \bfE[\frac{ \bfE(e_{u,i}^2|x_{u,i}) }{ p_{u,i} }  ]  -  \Big \{ \bfE( e_{u,i}) \Big \}^2 \Biggr ] \\ 
        ={}&  |\cD|^{-1} \cdot  \Biggl [  \bfE[\frac{ \sigma^2(x_{u,i}) + g_{u,i}^2 }{ p_{u,i} }  ]  -  \Big \{ \bfE( e_{u,i}) \Big \}^2 \Biggr ],
    \end{align*}
where the third equation follows by the law of iterated expectations and the unconfoundedness assumption. 
The variance of DR estimator is derived by   
    \begin{align*}
    |\cD| \cdot \V(\cL_{DR}) =
    {}&  \V \left ( g_{u,i} + \frac{ o_{u,i} ( e_{u,i} - g_{u,i} ) }{ p_{u,i} }  \right )      \\   
    ={}& \V \left  ( e_{u,i} + \frac{ o_{u,i} - p_{u,i} }{ p_{u,i} } ( e_{u,i} - g_{u,i} )  \right )    \\  
    ={}&  \V( e_{u,i})  + \V \left ( \frac{ o_{u,i} - p_{u,i} }{ p_{u,i} } ( e_{u,i} - g_{u,i} )   \right ) \\
        ={}&  \bfE[ e_{u,i}^2 ] - [\bfE(e_{u,i})]^2  + \V \left ( \frac{ o_{u,i} - p_{u,i} }{ p_{u,i} } ( e_{u,i} - g_{u,i} )  \right ) \\
    ={}&  \bfE[\sigma^2(x_{u,i}) +  g_{u,i}^2]  - [\bfE(e_{u,i})]^2 +
            \bfE \left  ( \frac{ (o_{u,i} - p_{u,i})^2 }{ p^2_{u,i} } ( e_{u,i} - g_{u,i} )^2  \right ) \\
    ={}&  \bfE[\sigma^2(x_{u,i}) +  g_{u,i}^2]  - [\bfE(e_{u,i})]^2 + \bfE \left  ( \frac{ \bfE\{(o_{u,i} - p_{u,i})^2 | x_{u,i}\} }{ p^2_{u,i} } \cdot \bfE\{( e_{u,i} - g_{u,i} )^2 | x_{u,i} \} \right )  \\ 
    ={}& \bfE[\sigma^2(x_{u,i}) +  g_{u,i}^2]  - [\bfE(e_{u,i})]^2 +
            \bfE \left ( \frac{ p_{u,i}(1-p_{u,i}) \sigma^2(x_{u,i}) }{ p^2_{u,i} } \right )    \\
    ={}& \bfE[ \frac{\sigma^2(x_{u,i}) }{p_{u,i}} + g_{u,i}^2 ]  - [\bfE(e_{u,i})]^2,        
    \end{align*}        
where the fifth equation holds by noting that 
    \[  \bfE \left [ e_{u,i} \frac{(o_{u,i} - p_{u,i}) }{ p_{u,i} } ( e_{u,i} - g_{u,i} )  \right]  = \bfE \left [  \frac{ \bfE(o_{u,i} - p_{u,i} | x_{u,i}) }{ p_{u,i} } \cdot \bfE\{ e_{u,i}( e_{u,i} - g_{u,i} ) | x_{u,i}\}  \right]   =   0.   \]    

Since $\bfE[ o_{u,i} e_{u,i} + (1-o_{u,i}) g_{u,i} ] = \bfE[ g_{u,i} ] = \bfE[ e_{u,i} ]$, we have 
    \begin{align*}
         |\cD| \cdot  \V( \cL_{EIB} ) ={}& \V \left ( o_{u,i} e_{u,i} + (1-o_{u,i}) g_{u,i} \right ) \\
         ={}& \bfE \Big [ \{ o_{u,i} e_{u,i} + (1-o_{u,i}) g_{u,i} \}^2\Big ] - [\bfE(e_{u,i})]^2 \\
         ={}&  \bfE \Big [ o_{u,i} e^2_{u,i} + (1-o_{u,i}) g^2_{u,i} \Big ] - [\bfE(e_{u,i})]^2  \\
         ={}& \bfE \Big [ p_{u,i}\{ \sigma^2(x_{u,i}) +  g_{u,i}^2 \}  + (1-p_{u,i}) g^2_{u,i} \Big ] - [\bfE(e_{u,i})]^2 \\
         ={}& \bfE \Big [ p_{u,i} \sigma^2(x_{u,i}) +  g_{u,i}^2 \Big ] - [\bfE(e_{u,i})]^2 
    \end{align*}    
\end{proof}

\section{Proof of Theorem 1}  \label{app-tmp}

   \begin{proof}[Proof of Theorem 1]  
The parameter $\eta$ is solved by minimizing 
		\[ \sum_{ (u,i) \in \cD } o_{u,i} \cdot \biggl \{  e_{u,i} -   \hat e_{u,i}  - \eta (\frac{1}{ \hat p_{u,i} } - 1 )  \biggr  \}^2.
		  \]   
Taking the first derivative of the above loss with respect to $\eta$ and setting it to zero leads to that  
		\begin{equation}\label{eq1-appendix} \sum_{ (u,i) \in \cD } o_{u,i} \cdot \biggl \{  e_{u,i} -  \hat e_{u,i}  - \eta (\frac{1}{ \hat p_{u,i} } - 1 ) \biggr  \}   \cdot  (1/ \hat p_{u,i} - 1)  = 0,
		  \end{equation}
	which implies that 
			\begin{equation*} \sum_{ (u,i) \in \cD } o_{u,i} \cdot \left \{  e_{u,i} -  \tilde e_{u,i} \right  \}   \cdot  (1/ \hat p_{u,i} - 1)  = 0,
		  \end{equation*}
namely, the equation (\ref{eq5}) holds. This finishes the proof of Theorem 1(a).    If $\hat e_{u,i}$ already satisfies equation \ref{eq5}), then $\eta = 0$ is a solution of equation (\ref{eq1-appendix}). Let $\hat \eta$ is another solution of equation (\ref{eq1-appendix}). Since  the solution of  equation (\ref{eq1-appendix}) is unique, then $\hat \eta$ will converges to 0.  This proves the conclusion of Theorem 1(b).

\end{proof}

\section{Proof of Theorem 2}  \label{app-proof2}

      \begin{proof}[Proof of Theorem 2] 
   
The result of Theorem 2(a) is obvious. To show Theorem 2(b). We first claim that   
if $\hat e_{u,i}$ is an accurate estimate of $g_{u,i}$, i.e., $\hat e_{u,i} = g_{u,i}$, then it will satisfy equation (\ref{eq5}). It holds immediately from the following calculations 
    \begin{align*}
     & \frac{1}{|\cD|} \sum_{(u,i) \in \cD } o_{u,i} \{  e_{u,i} -  \hat e_{u,i}  \}  \cdot  ( \frac{1}  {p_{u,i}} - 1)  \\
    ={}& \frac{1}{|\cD|} \sum_{(u,i) \in \cD } o_{u,i} \{  e_{u,i} -  g_{u,i}  \}  \cdot  ( \frac{1}  {p_{u,i}} - 1)  \\
    ={}& \bfE \Big[ o_{u,i} \{  e_{u,i} -  g_{u,i}  \}  \cdot  ( \frac{1}  {p_{u,i}} - 1)     \Big ]\\
    ={}&  \bfE \Big[ \bfE(o_{u,i}|x_{u,i})  \cdot \bfE\{ e_{u,i} -  g_{u,i} | x_{u,i} \}  \cdot  ( \frac{1}  {p_{u,i}} - 1)     \Big ] \\
    ={}&  0. 
    \end{align*}
Thus,  if $\hat e_{u,i}$ not satisfy equation (\ref{eq5}), then $\hat e_{u,i} \neq g_{u,i}$. Given $\hat e_{u,i}$ and $\tilde e_{u,i}$, the bias of $\cL_{EIB}$ is 
    \begin{align*}
    \text{Bias}(\cL_{EIB}) ={}&  \bfE[ o_{u,i} e_{u,i} + (1-o_{u,i}) \hat e_{u,i}  ] -  \bfE[ e_{u,i} ]  \\
    ={}&  \bfE[ (1-o_{u,i}) (  \hat e_{u,i}  - e_{u,i} )  ] \\
    ={}&  \bfE[ (1-p_{u,i}) ( \hat e_{u,i}  - g_{u,i} ) ],  
    \end{align*}
and the bias of $\cL_{TDR}$ is 
    \begin{align*}
    \text{Bias}(\cL_{TDR}) 
    ={}& \bfE \left  ( e_{u,i} + \frac{ (o_{u,i} - p_{u,i}) }{ p_{u,i} } ( e_{u,i} - \tilde e_{u,i} )  \right )  - \bfE[ e_{u,i} ]  \\ 
    ={}& \bfE \left  ( \frac{ \bfE(o_{u,i} - p_{u,i} | x_{u,i}) }{ p_{u,i} } \bfE\{ e_{u,i} - \tilde e_{u,i} | x_{u,i} \}  \right ) \\
    ={}& \bfE \left  ( \frac{ 0 }{ p_{u,i} } \cdot (g_{u,i} - \tilde e_{u,i})  \right ) \\
    ={}& 0. 
    \end{align*}
This proves the result of Theorem 2(b). 

    \end{proof}

 \begin{proof}[Biases of DR and TDR]   Given $\hat p_{u,i}$ and $\tilde e_{u,i}$ for all $(u,i) \in \cD$, the bias of TDR is 
\begin{align*}
    \text{Bias}(\cL_{TDR}) 
    ={}& \bfE \left  ( e_{u,i} + \frac{ (o_{u,i} - \hat p_{u,i}) }{ \hat p_{u,i} } ( e_{u,i} - \tilde e_{u,i} )  \right )  - \bfE[ e_{u,i} ]  \\ 
    ={}& \bfE \left  ( \frac{ \bfE(o_{u,i} - \hat p_{u,i} | x_{u,i}) }{ \hat p_{u,i} } \bfE\{ e_{u,i} - \tilde e_{u,i} | x_{u,i} \}  \right ) \\
    ={}& \bfE \left  ( \frac{ (p_{u,i} -  \hat p_{u,i})  }{\hat p_{u,i} } \cdot (g_{u,i} - \tilde e_{u,i})  \right ).  
    \end{align*}

Similarly, given $\hat p_{u,i}$ and $\hat e_{u,i}$ for all $(u,i) \in \cD$, the bias of DR is  
\begin{align*}
  \text{Bias}(\cL_{DR}^{(0)}) 
    ={}& \bfE \left  ( \frac{ (p_{u,i} -  \hat p_{u,i})  }{\hat p_{u,i} } \cdot (g_{u,i} - \hat e_{u,i})  \right ).  
    \end{align*}
\end{proof}

\section{Real-world Experimental Protocols and Details}
\label{app-real}
{\bf Experimental protocols and details.} For real-world experiments, the following four metrics were considered as the evaluation metrics: \emph{MSE, AUC, NDCG@5,} and \emph{NDCG@10}. For fast convergence in the learning phase, Adam is utilized as the optimizer for all models. We tune the learning rate in $\{0.001,0.005,0.01,0.05,0.1\}$, weight decay in $[1\mathrm{e}-6, 1\mathrm{e}-2]$ at $10 \mathrm{x}$ multiplicative ratio, and batch size in $\{128,256,512,1024,2048\}$ for {\bf Coat} and $\{1024, 2048, 4096, 8192, 16384\}$ for {\bf Music! R3}. Specifically for the propensity training, we tune the clipping threshold in $\{0.05, 0.10, 0.15, 0.20\}$. After finding out the best configuration on the validation set, we evaluate the trained models on the MAR test set. 
Experiments are conducted using NVIDIA GeForce RTX 3090.

\end{document}